\documentclass[twocolumn]{aastex631}

\usepackage{booktabs}
\usepackage{tabularx}
\usepackage{array}
\usepackage{comment}
\usepackage{enumitem}

\DeclareRobustCommand{\mytt}[1]{%
  \ifmmode
    \text{\texttt{\detokenize{#1}}}%
  \else
    \texttt{\detokenize{#1}}%
  \fi
}
\received{2026 April 9}
\revised{2026 August 31}
\accepted{2026 September 21}

\begin{document}

\title{J-PLUS: Identifying Outlier Spectral Energy Distributions in the Fourth Data Release Using Unsupervised Machine Learning}

\author[0000-0001-8741-8642]{F. Quispe-Huaynasi}
\affiliation{Departamento de Física, Universidade Federal do Espírito Santo, Vitória, ES 29075-910, Brazil.}
\affiliation{Observatório Nacional, Rio de Janeiro, RJ 20921-400, Brazil.}

\author[0000-0001-7059-5116]{F. Roig}
\affiliation{Observatório Nacional, Rio de Janeiro, RJ 20921-400, Brazil.}

\author[0000-0003-0506-8269]{V. Loaiza-Tacuri}
\affiliation{Departamento de Física, Universidade Federal de Sergipe, São Cristóvão, SE 49100-000, Brazil.}

\author[0000-0003-4479-1265]{V. M. Placco}
\affiliation{NSF NOIRLab, Tucson, AZ 85719, USA.}

\author[0000-0002-7773-1579]{V. Marra}
\affiliation{Departamento de Física, Universidade Federal de Ouro Preto, 35400-000, Ouro Preto, MG, Brazil}
\affiliation{INAF – Osservatorio Astronomico di Trieste, via Tiepolo 11, 34131 Trieste, Italy}
\affiliation{IFPU – Institute for Fundamental Physics of the Universe, via Beirut 2, 34151, Trieste, Italy}

\author{F. Jiménez-Esteban}
\affiliation{Centro de Astrobiolog\'{\i}a, CSIC-INTA, Camino bajo del castillo s/n, E-28692, Villanueva de la Can\~ada, Madrid, Spain.}

\author[0000-0003-4922-5131]{A. del Pino}
\affiliation{Instituto de Astrofísica de Andalucía (IAA-CSIC), Glorieta de la Astronomía S/N, 18008, Granada, Spain.}

\author[0000-0001-5261-4336]{B. Anguiano}
\affiliation{Centro de Estudios de Física del Cosmos de Aragón (CEFCA), 44001, Teruel, Spain.}

\author[0000-0002-8351-8854]{J. A. Alzate-Trujillo}
\affiliation{Centro de Estudios de Física del Cosmos de Aragón (CEFCA), 44001, Teruel, Spain.}

\author[0000-0001-9205-2307]{S. Daflon}
\affiliation{Observatório Nacional, Rio de Janeiro, RJ 20921-400, Brazil.}

\author{C. López-Sanjuan}
\affiliation{Centro de Estudios de Física del Cosmos de Aragón (CEFCA), 44001, Teruel, Spain.}
\affiliation{Unidad Asociada CEFCA-IAA, CEFCA, Unidad Asociada al CSIC por el IAA y el IFCA, 44001, Teruel, Spain.}

\author[0000-0003-1793-200X]{P. Cruz}
\affiliation{Centro de Astrobiolog\'{\i}a, CSIC-INTA, Camino bajo del castillo s/n, E-28692, Villanueva de la Can\~ada, Madrid, Spain.}

\author[0000-0002-5045-9675]{A. Alvarez-Candal}
\affiliation{Instituto de Astrofísica de Andalucía (IAA-CSIC), Glorieta de la Astronomía S/N, 18008, Granada, Spain.}

\author[0000-0001-6987-1531]{E. Machado-Pereira}
\affiliation{Observatório Nacional, Rio de Janeiro, RJ 20921-400, Brazil.}

\author{J. Alcaniz}
\affiliation{Observatório Nacional, Rio de Janeiro, RJ 20921-400, Brazil.}

\author{R. E. Angulo}
\affiliation{Donostia International Physics Center, 20018, Donostia-San Sebastián, Spain.}

\author{J. Cenarro}
\affiliation{Centro de Estudios de Física del Cosmos de Aragón (CEFCA), 44001, Teruel, Spain.}
\affiliation{Unidad Asociada CEFCA-IAA, CEFCA, Unidad Asociada al CSIC por el IAA y el IFCA, 44001, Teruel, Spain.}

\author{D. Cristóbal-Hornillos}
\affiliation{Centro de Estudios de Física del Cosmos de Aragón (CEFCA), 44001, Teruel, Spain.}

\author{R. Dupke}
\affiliation{Observatório Nacional, Rio de Janeiro, RJ 20921-400, Brazil.}

\author{A. Ederoclite}
\affiliation{Centro de Estudios de Física del Cosmos de Aragón (CEFCA), 44001, Teruel, Spain.}

\author{A. Hernán-Caballero}
\affiliation{Centro de Estudios de Física del Cosmos de Aragón (CEFCA), 44001, Teruel, Spain.}
\affiliation{Unidad Asociada CEFCA-IAA, CEFCA, Unidad Asociada al CSIC por el IAA y el IFCA, 44001, Teruel, Spain.}

\author{C. Hernández-Monteagudo}
\affiliation{Centro de Estudios de Física del Cosmos de Aragón (CEFCA), 44001, Teruel, Spain.}

\author[0000-0002-9026-3933]{A. Marin-Franch}
\affiliation{Centro de Estudios de Física del Cosmos de Aragón (CEFCA), 44001, Teruel, Spain.}
\affiliation{Unidad Asociada CEFCA-IAA, CEFCA, Unidad Asociada al CSIC por el IAA y el IFCA, 44001, Teruel, Spain.}

\author{M. Moles}
\affiliation{Centro de Estudios de Física del Cosmos de Aragón (CEFCA), 44001, Teruel, Spain.}

\author{L. Sodré Jr.}
\affiliation{Instituto de Astronomia, Geofísica e Ciências Atmosféricas, Universidade de São Paulo, SP 05508-090, Brazil.}

\author{J. Varela}
\affiliation{Centro de Estudios de Física del Cosmos de Aragón (CEFCA), 44001, Teruel, Spain.}

\author{H. Vázquez-Ramió}
\affiliation{Centro de Estudios de Física del Cosmos de Aragón (CEFCA), 44001, Teruel, Spain.}
\affiliation{Unidad Asociada CEFCA-IAA, CEFCA, Unidad Asociada al CSIC por el IAA y el IFCA, 44001, Teruel, Spain.}

\author{J. Zaragoza-Cardiel}
\affiliation{Centro de Estudios de Física del Cosmos de Aragón (CEFCA), 44001, Teruel, Spain.}
\affiliation{Unidad Asociada CEFCA-IAA, CEFCA, Unidad Asociada al CSIC por el IAA y el IFCA, 44001, Teruel, Spain.}


\begin{abstract}
The selection of candidates belonging to rare or underrepresented stellar populations, as well as novel objects, is essential for spectroscopic follow-up. In this work, we demonstrate how combining broad- and narrow-band photometric information with machine-learning techniques enables this selection. Using photometric data in twelve filters from the Javalambre Photometric Local Universe Survey (J-PLUS) as input, we applied an anomaly detection technique based on autoencoders to identify spectral energy distributions (SEDs) that deviate from the dominant pattern in the J-PLUS objects sample. Subsequently, we applied the t-distributed stochastic neighbor embedding (t-SNE) algorithm to the J-PLUS colors of the outlier SEDs population. Combining information from the SIMBAD database and other astronomical surveys to classify the outliers identified by the autoencoder, we are able to characterize over-densities in the t-SNE map. Our study specifically focuses on well-defined islands in the t-SNE space, corresponding to specific populations. We identified a large sample of white dwarfs and low-mass stars, including a small subset of polluted white dwarfs, and report potential candidates for spectroscopic follow-up, including 19 carbon-rich stars, 11 cataclysmic variables and 170 white-dwarfs-main-sequence binaries, all previously unreported in the literature.

\end{abstract}

\keywords{Chemically peculiar stars --- Carbon stars --- White dwarf stars --- Low mass stars --- Binary stars --- Cataclysmic variable stars}

\section{Introduction}\label{sec:intro}

The selection of reliable stellar population candidates, either from low-resolution spectroscopy or multi-band photometric surveys, is a fundamental task in astronomy. This is especially relevant for potential high-resolution spectroscopic follow-up using either ground- or space-based facilities. Also, identifying rare or underrepresented stellar populations is essential for uncovering new astrophysical phenomena and constraining models of stellar and galactic evolution. Traditionally, candidate selection has relied on color-magnitude and color-color diagrams, in which common stellar and extragalactic populations occupy characteristic loci determined by their spectral energy distributions \citep{2000AJ....120.2615F,2001AJ....122.1861S}. Broad-band surveys such as the Sloan Digital Sky Survey \citep[SDSS;][]{2000AJ....120.1579Y} demonstrated the effectiveness of this approach. For example, \citet{2004ApJ...615L.141S} mapped the distribution of main-sequence stars, quasars, and white dwarfs in the $(g-r)$ versus $(u-g)$ plane, while \citet{2002AJ....123..430S} showed that cataclysmic variables occupy distinct regions of color-color space owing to their accretion-dominated spectral energy distributions. Nevertheless, many rare populations overlap the loci of common sources or exhibit only subtle deviations in individual color projections, limiting the effectiveness of manually defined color cuts.

A significant advance toward systematic anomaly detection in photometric surveys was made by \citet{2007AJ....134.2398C}. They modeled the stellar locus in seven-dimensional color space using SDSS and 2MASS photometry and introduced the seven-dimensional color distance (7DCD): a metric that quantifies an object's separation from the stellar locus. By exploiting the joint distribution of multiple colors, the 7DCD enabled the identification of photometric outliers that are not readily apparent in individual color-color diagrams. The resulting outlier sample was enriched in unusual populations, including white dwarf-M dwarf binaries, quasars, and M giant-carbon star candidates, demonstrating the potential of multidimensional photometric information for discovering rare sources.

More recently, the increasing volume and complexity of astronomical surveys have motivated the development of machine learning approaches for anomaly detection. Rather than relying exclusively on manually defined selection criteria or explicit parametric descriptions of the data distribution, these methods learn patterns directly from observations and identify deviations in high-dimensional parameter spaces. Examples include one-class support vector machines applied to WISE photometric catalogs \citep{2017A&A...606A..39S}, unsupervised approaches for detecting anomalous light curves \citep{2019MNRAS.484..834G}, and deep representation learning methods based on autoencoders for identifying spectroscopic outliers in DESI data \citep{2023ApJ...956L...6L} and low-resolution SDSS spectra \citep{2019BSRSL..88..174S}.

In parallel, several photometric surveys have expanded beyond traditional broad-band systems by incorporating medium- and narrow-band filters, aiming to improve redshift measurements, enhance stellar population discrimination, and enable more precise stellar parameter determination. These include: the Physics of the Accelerating Universe Survey \citep[PAUS;][]{2019MNRAS.484.4200E}, SkyMapper \citep{2019MNRAS.489.5900D}, the Pristine survey \citep{2017MNRAS.471.2587S}, the Javalambre Photometric Local Universe Survey \citep[J-PLUS;][]{2019A&A...622A.176C}, the Southern Photometric Local Universe Survey \citep[S-PLUS;][]{2019MNRAS.489..241M}, the miniJPAS \citep{2021A&A...653A..31B}, and the Javalambre Physics of the Accelerating Universe Astrophysical Survey \citep[J-PAS;][]{2014arXiv1403.5237B}. While SkyMapper and Pristine include filters designed to probe the Ca II H\&K absorption feature and identify metal-poor stars, J-PLUS, S-PLUS, and J-PAS employ multiple medium- and narrow-band filters centered on key spectral features sensitive to stellar properties such as effective temperature, metallicity, and surface gravity. These enhanced photometric spectral energy distributions provide a richer representation of stellar populations than conventional broad-band colors, and have enabled stellar classification and parameter estimation through both model-based and machine-learning approaches \citep{2019A&A...622A.176C, 2019A&A...622A.182W, 2021ApJ...912..147W, 2022A&A...657A..35G, 2024MNRAS.527.6173Q, 2024ApJ...974..192H, 2025A&A...693A.306F}.

In this work, we apply an unsupervised anomaly detection technique based on a deep autoencoder to identify spectral energy distributions (SEDs) in J-PLUS that deviate from the dominant stellar population. The autoencoder is trained to compress and reconstruct the twelve-dimensional photometry of typical stars; then, objects with the highest reconstruction error are flagged as candidate outliers. This approach was previously demonstrated on S-PLUS data by \citet{2025AJ....169..332Q}, and we follow here the methodology described in that paper. In this paper, we extend that analysis to J-PLUS, exploiting the fact that the two surveys share an identical filter system but cover different sky regions. Thus, the autoencoder may reveal distinct populations of interest in the J-PLUS footprint.

To inspect the candidate outliers, we project the high-residual objects into two dimensions using t-distributed stochastic neighbor embedding (t-SNE), which reveals clustering structure among the anomalies and guides visual inspection. We then cross-match these candidates with external astronomical databases to assess their nature and select promising targets for high-resolution spectroscopic follow-up. It is worth emphasizing that objects classified as outliers are not necessarily intrinsically peculiar or rare in an astronomical sense, although some may indeed be; instead, they correspond to objects that are underrepresented or absent in the training distribution.

This paper is organized as follows. The J-PLUS data are briefly described in Sect.~\ref{sec:jplus_data}. Sect.~\ref{sec:three} presents the autoencoder methodology used to identify outlier SEDs, including the training procedure, network architecture, and its application to the J-PLUS data. The application of t-SNE to the anomalous sample, aimed at segregating different populations, is presented in Sect.~\ref{sec:fourth}. The identification, selection, and description of some populations identified in t-SNE space are presented in Sect.~\ref{sec:five}. Finally, Sect.~\ref{sec:six} summarizes the results, and presents the discussion and conclusions.

\section{The J-PLUS data} \label{sec:jplus_data}

The magnitudes used as input to train the autoencoder, and the colors used as input for the t-SNE algorithm, came from the 12 photometric bands of the J-PLUS survey \citep{2019A&A...622A.176C}. The J-PLUS photometric system combines SDSS broad-band filters (\textit{u}, \textit{g}, \textit{r}, \textit{i}, and \textit{z}), medium-band filters (J0395, J0410, J0430, J0515, and J0861), and narrow-band filters (J0378 and J0660), that cover wavelengths from 3500~$\mathrm{\AA}$ to 10,000~$\mathrm{\AA}$. This combination of filters enables the construction of reliable stellar SEDs. The medium- and narrow-band filters are strategically positioned to sample spectral regions containing absorption and emission features sensitive to stellar parameters and the properties of specific stellar populations, including H$\alpha$, H$\delta$, the Ca$\,$\textsc{ii} H\&K doublet, the Mg$\,b$ triplet, the Ca$\,$\textsc{ii} triplet, among others \citep{2019A&A...622A.176C}. The central wavelengths, FWHMs, and associated spectral features of the individual filters are listed in Table~\ref{tab:jplus_filters}. 

Although the filters do not necessarily resolve individual spectral features, their integrated fluxes retain information about the underlying stellar spectra. Recent studies \citep{2019A&A...622A.182W, 2022A&A...664A..38W, 2022A&A...659A.181Y, 2024MNRAS.527.6173Q, 2025A&A...693A.306F} have highlighted the potential of machine-learning methods to exploit these multivariate and nonlinear relationships, mitigating the parameter degeneracies by learning the mapping between photometric measurements and stellar properties from spectroscopic characterized samples. Applications to J-PLUS and S-PLUS photometry have demonstrated their effectiveness in estimating stellar parameters and chemical abundances, including the identification of metal-poor stars \citep{2022A&A...657A..35G} and carbon enhanced metal-poor stars \citep{2021ApJ...912..147W}. Independently, color-color diagrams with narrow-band photometry have also been applied to select metal-poor stars for spectroscopic follow-up \citep{2022ApJS..262....8P}, demonstrating the diagnostic power of these filters for identifying chemically distinct stellar populations.

The J-PLUS is being conducted at the Observatorio Astrofísico de Javalambre \citep[OAJ;][]{oaj} with the 83 cm Javalambre Auxiliary Survey Telescope (JAST80) and the $9200 \times 9200$ pixel camera T80Cam \citep{t80cam}, providing a $2\,\deg^2$ field of view per pointing. The J-PLUS fourth data release\footnote{\url{https://www.j-plus.es/datareleases/data_release_dr4}} (DR4) used in this work comprises 2,563 fields, covering approximately 5,000 $\deg^2$ and containing about 47.3 million objects with magnitude $r \leq 21$.

\begin{table}
\centering
\caption{Characteristics of the J-PLUS photometric bands, including central wavelength, FWHM, and associated spectral features.}
\label{tab:jplus_filters}
\setlength{\tabcolsep}{10pt}
\begin{tabularx}{\columnwidth}{lccc}
\hline 
Filter & $\lambda_{\rm c}$ & FWHM & Spectral features \\       &          \AA       &   \AA  &                   \\
\hline
uJAVA  &   3485  & 508  &  Balmer-break region           \\
J0378  &   3785  & 168  &  [OII]                         \\
J0395  &   3950  & 100  &  Ca\,H + K                     \\
J0410  &   4100  & 200  &  $\mathrm{H}_{\delta}$         \\
J0430  &   4300  & 200  &  G band                        \\
gSDSS  &   4803  & 1409 &  SDSS-like g                   \\
J0515  &   5150  & 200  &  Mg \textit{b} triplet         \\
rSDSS  &   6254  & 1388 &  SDSS-like r                   \\
J0660  &   6600  & 138  &  $\mathrm{H}_{\alpha}$         \\
iSDSS  &   7668  & 1535 &  SDSS-like i                   \\
J0861  &   8610  & 400  &  Ca triplet                    \\
zSDSS  &   9114  & 1409 &  SDSS-like z                   \\
\hline
\end{tabularx}
\end{table}

\section{Outliers identification using autoencoder}\label{sec:three}
Outlier detection methods can be divided into statistical and machine-learning approaches \citep{10.1145/1541880.1541882}. The choice of method depends on the size of the dataset and its characteristics, such as feature correlation and data dimensionality \citep{10.1145/3439950, Aggarwal2017}. In this work, we adopted an undercomplete vanilla autoencoder architecture, which has been used for anomaly detection \citep{10.1145/2689746.2689747}. Previous studies have reported that autoencoder-based approaches can outperform traditional distance- and density-based methods in specific scenarios involving high-dimensional and structured data \citep{10.1145/3097983.3098052, 9347460}. 

Autoencoders are classified as unsupervised (or self-supervised) neural networks, and are composed of three main components: an encoder, a latent space, and a decoder. The encoder maps the input data into a latent representation. When the dimensionality of this latent space is lower than that of the input data, the model is referred to as an undercomplete autoencoder. The decoder maps the latent representation to an output that lies in the same space as the input data. During training, the encoder–decoder network is optimized to reconstruct the input data by minimizing a reconstruction loss. 

Vanilla autoencoders and their extensions have been widely applied to dimensionality reduction tasks \citep{2006Sci...313..504H, 2015MNRAS.452..158Y, 2020AJ....160...45P} and as generative models \citep{8354080, NAZABAL2020107501}. Autoencoder-based architectures have also been successfully applied to a wide range of astronomical problems \citep{2020AJ....160...45P, 2021AAS...23810305C, 2023arXiv231216687G, 2025ApJ...979....5L}. 

Since we aimed to identify outlier SEDs in the J-PLUS photometric data, we used the autoencoder as an anomaly detector, flagging data instances that do not follow the expected pattern of the majority of the data, or that deviate significantly from the typical data \citep{10.1145/3439950}. These anomalous data represent a small fraction of the total sample. The idea is to train the autoencoder on a sample of normal or non-anomalous SEDs (majority class), allowing it to learn the typical patterns of the normal data. After training, the reconstruction error serves as an anomaly score, being expected to be lower for normal data compared to anomalous data. 

For a given source, the reconstruction error was defined by the mean square error (MSE) as:
\begin{equation}
  E_{\mathrm{rec}}=\frac{1}{N}\sum_{j=1}^{N}\left(m_{j}^{\mathrm{obs}}-m_{j}^{\mathrm{rec}}\right)^{2} \label{eq:err_rec}
\end{equation}
where $N=12$ is the number of filters, $m_{j}^{\mathrm{obs}}$ is the observed magnitude in a given filter, and $m_{j}^{\mathrm{rec}}$ is the reconstructed magnitude value in the same filter.
An alternative definition of $E_\mathrm{rec}$ can be obtained using the $\chi^2$ instead of the MSE, which would allow to account for the effect of the individual errors in each filter. Here, however, the effect of the individual errors was partially mitigated by imposing an upper cutoff in the errors, as we describe in Sect.~\ref{sec:train_val_test_samples}. Further discussion on the effects of not including the observational errors in the training of the autoencoder is given in Sect.~\ref{sec:obs-errors}.

For more in depth details on the methodology applied here, we refer the reader to \cite{2025AJ....169..332Q}.

\subsection{Autoencoder setup}\label{subsection:autoencoder_architecture}

The autoencoder architecture was implemented using \texttt{keras} \citep{chollet2015keras}, and the training settings were optimized with \texttt{KerasTuner} \citep{omalley2019kerastuner} using Bayesian optimization over 30 trials. The search space included: the bottleneck size (2--6), the number of encoder/decoder layers (2--4), the learning rate ($10^{-4}$--$10^{-2}$, log-uniform), the L1 ($10^{-6}$--$10^{-4}$) and L2 ($10^{-5}$--$10^{-3}$) regularization strengths, the first-layer ratio (0.5--0.8), and the compression rate (0.5--0.75). Each trial was trained for up to 50 epochs with a batch size of 256 using the Adam optimizer with gradient clipping. The objective function was the validation loss, with early stopping (patience 10) to restore the best weights and reduce the learning rate on plateau (patience 5, factor 0.5).

The resulting final architecture was a symmetric autoencoder with 12 input features, two encoder layers of 9 and 6 units, a bottleneck of 5 units, two decoder layers of 6 and 9 units, and a linear output layer of 12 units. Hidden layers used ReLU activations, while the bottleneck and output used linear activations. The encoder and decoder dense layers were regularized with a combined L1/L2 penalty. Further discussion on the autoencoder architecture is given in Sect.~\ref{sec:auto-arch}.

\subsection{Training, Validation, and Test Samples}\label{sec:train_val_test_samples}

\begin{figure}[ht]
\centering
\includegraphics[width=\linewidth]{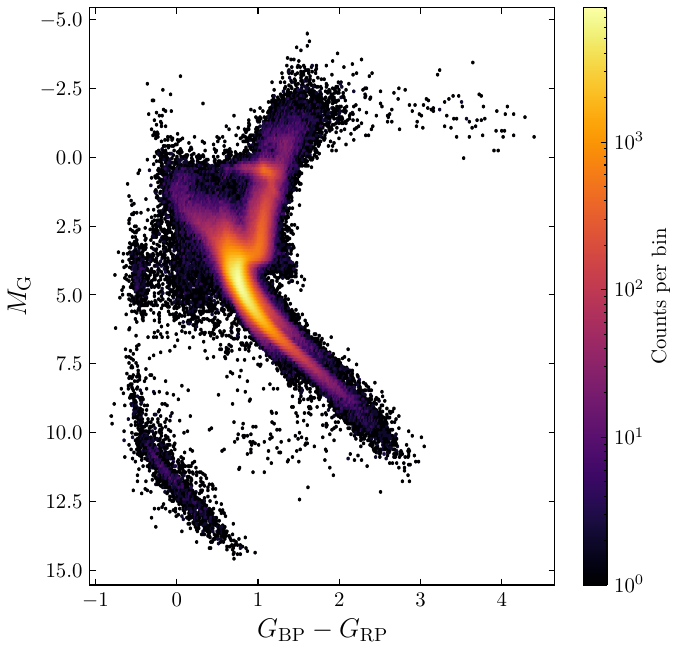}
\caption{Color–magnitude diagram, colored by point density, of the sample of $\sim 2.7$ million stars used to develop the autoencoder model. The horizontal axis shows the \textit{Gaia} color $G_\mathrm{BP}-G_\mathrm{RP}$, and the vertical axis show the absolute G-band magnitude, $M_\mathrm{G}$, calculated from the apparent G magnitude and the photo-geometric distance provided by \citet{2021AJ....161..147B}.}
\label{fig:cmd_training_sample}
\end{figure}

Our parent sample was constituted by J-PLUS DR4 sources with complete photometric measurements in the 12 filters. We selected sources with magnitudes smaller than 21 in all the 12 filters, together with magnitude uncertainties smaller than 0.15. We further applied the following selection cuts:
\begin{enumerate}[label=\alph*), leftmargin=*]
    \item Based on the \texttt{SExtractor} \citep{2010ascl.soft10064B} classification, we required $\mathtt{class\_star} > 0.95$.
    \item Based on the \textit{Gaia} flags, we required: (i) positive parallax ($\mathtt{\varpi > 0}$); (ii) $\mathtt{RUWE < 1.2}$, to ensure a good astrometric solution and minimize contamination from unresolved binaries or other sources with poor astrometric solutions\footnote{$\mathtt{RUWE}$ is the Re-normalised Unit Weight Error}; (iii) $\mathtt{in\_qso\_candidates = 0}$ and $\mathtt{in\_galaxy\_candidates = 0}$, to remove probable quasar and galaxy candidates, respectively; (iv) $\mathtt{non\_single\_star = 0}$, to exclude sources identified as non-single stars, such as astrometric or spectroscopic binaries and eclipsing binaries; and (v) $\mathtt{duplicated\_source = 0}$, to remove duplicated sources.
\end{enumerate}

These selection cuts resulted in a parent sample of $\sim 2.7$ million sources that was used for model development. Of these sources, 80\% were used for training and validation, while the remaining 20\% were reserved as an independent test set.

Figure~\ref{fig:cmd_training_sample} shows the color–magnitude diagram for the sources with available distances provided by \citet{2021AJ....161..147B}.

\subsection{Application to the J-PLUS data}\label{subsec:app}

The autoencoder was trained using the 12 J-PLUS magnitudes as input. We chose to use magnitudes because we were interested in reconstructing the spectral energy distributions in their original photometric representation. This representation also facilitates the visual inspection of the SED morphology. The input magnitudes were previously scaled with the \texttt{MinMaxScaler()} function from the \texttt{scikit-learn} package \citep{scikit-learn}, to map each feature to the [0,1] range. This scaling was applied because the features have well-defined finite ranges, and substantial outliers are not expected given the quality cuts adopted to construct the parent and training samples. After the training, the final autoencoder model was applied to the entire parent sample (see Sect.~\ref{sec:train_val_test_samples}) to compute the reconstruction error for each source.

\begin{figure}[ht]
\centering
\includegraphics[width=\linewidth]{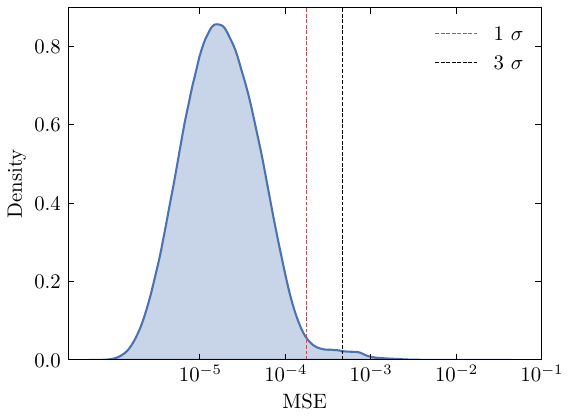}
\caption{Kernel Density Estimate (KDE) of the mean squared error (MSE) in $\log_{10}$ space. The $1\sigma$ threshold (red dashed line) was adopted to define the anomalous sample, while the $3\sigma$ threshold (black dashed line) is shown for reference.}
\label{fig:mse_error_distribution}
\end{figure}

We defined the outlier threshold as the mean reconstruction error plus $1\sigma$, with both the mean and $\sigma$ calculated from the reconstruction errors of the validation sample (see Fig. \ref{fig:mse_error_distribution}). This resulted in approximately 63,000 sources with reconstruction errors above this threshold and constitutes our outlier population. We adopted this relatively low threshold because a sensitivity analysis showed that increasing the threshold would preferentially remove scientifically interesting populations. In particular, among the populations discussed in this work (see Sect.~\ref{sec:five}), this effect is pronounced for low-mass stars: the 1$\sigma$ sample contains 16,698 candidates, of which 10,597 and 7,599 remain at 2$\sigma$ and 3$\sigma$, respectively. Of the 102 carbon star candidates identified at 1$\sigma$, only 34 and 11 remain at 2$\sigma$ and 3$\sigma$, respectively, meaning that a stricter threshold would remove 67\% and 89\% of this population. Regarding the polluted-white-dwarf candidates, only two of the five sources identified at 1$\sigma$ survive at both 2$\sigma$ and 3$\sigma$.

\begin{figure*}[ht]
\centering
\includegraphics[width=\linewidth]{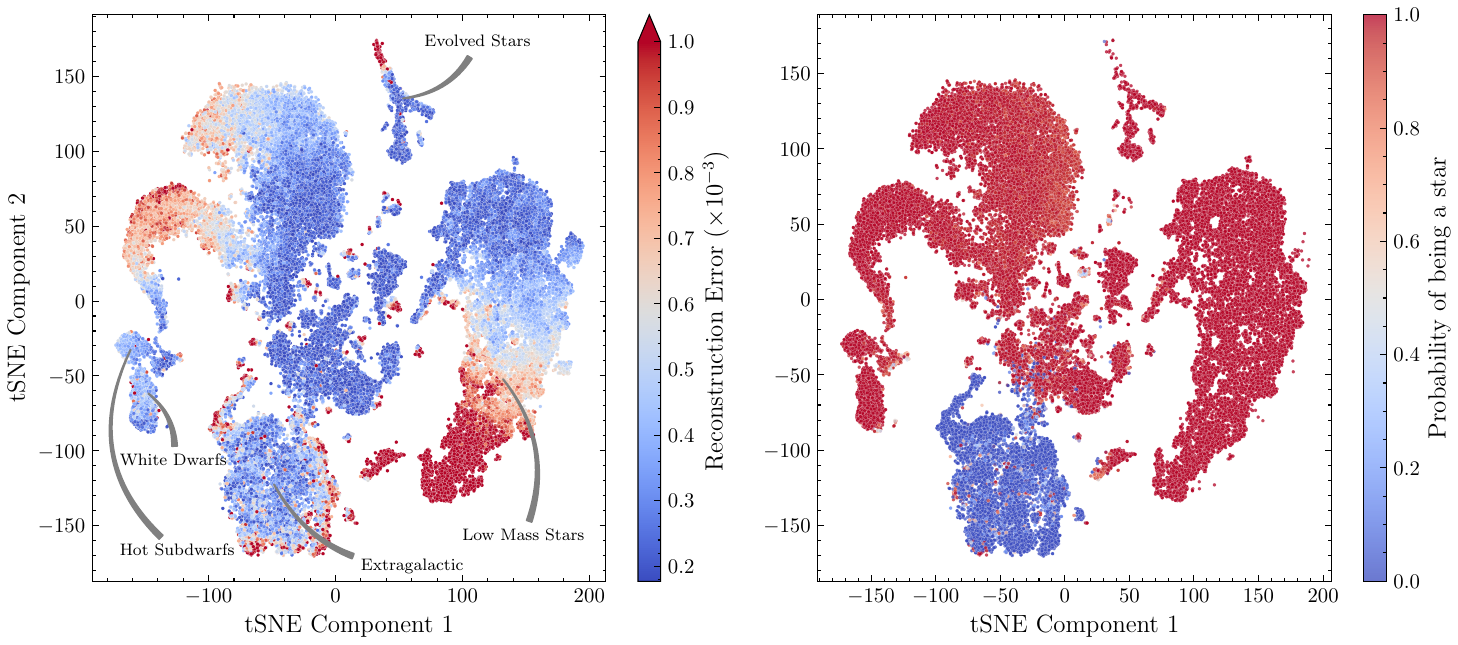}
\caption{Two-dimensional t-SNE maps. In the left panel, anomalous sources are colored by the reconstruction error, with sources showing large reconstruction errors highlighted in red. Here, the reconstruction error was computed from Eq. \ref{eq:err_rec}, but using the scaled values of $m_j$ (Sect.~\ref{subsec:app}). Using SIMBAD information, we also highlight five major groups in the map. In the right panel, sources are colored by the probability of being a star as provided by \mytt{LeMoNNADE} model. In both panels, except for the perplexity and the number of iterations, which were set to 50 and 5000, respectively, all the other hyperparameters were kept at their default values in the \mytt{scikit-learn} implementation of t-SNE.}
\label{fig:tsne_map}
\end{figure*}

In contrast, the cataclysmic-variable cluster is fully retained at both thresholds (51 of 51 members), while the WD–MS cluster remains essentially unaffected, with 303 members at 1$\sigma$, 303 at 2$\sigma$, and 300 at 3$\sigma$.

These results demonstrate that the $1\sigma$ threshold is important for retaining potentially interesting populations, particularly carbon star candidates, which are highly sensitive to the choice of threshold, whereas the CV and WD–MS populations are robust against changes in the threshold. It is worth noting that a 1$\sigma$ threshold corresponds to the $\sim$98th percentile of the sample, meaning that the anomalous population detected by the autoencoder corresponds to $\sim$2\% of the training population.

\section{Classification using \lowercase{t}-SNE}\label{sec:fourth}

To identify the types of sources flagged as anomalous by the autoencoder, we used the t-distributed Stochastic Neighbor Embedding algorithm \citep[t-SNE;][]{vanDerMaaten2008} to project the data into two dimensions for visual exploration of its structure. In high-dimensional spaces, distance metrics become less discriminative due to distance concentration, which reduces the effectiveness of density-based clustering methods when applied directly to the raw features in high dimensions. The low-dimensional embedding produced by t-SNE provides a more useful representation for visually identifying groups of similar objects. The t-SNE is a dimensionality reduction technique and has been applied in different areas of science, including astronomy \citep[][among others]{2017ApJS..228...24T, 2018A&A...619A.125A, 2019MNRAS.483.3196C}. The main idea of this non-linear algorithm is to place objects described by a high-dimensional vector (features) into a low-dimensional space (map), preserving the local neighbor property defined in the high-dimensional space through a probabilistic approach. Therefore, in our case, clumps in the t-SNE projection correspond to objects with similar colors (SEDs) and, given the physical nature of these observables, they are associated to sources sharing similar physical properties. 

\begin{figure}[ht]
\centering
\includegraphics[width=\linewidth]{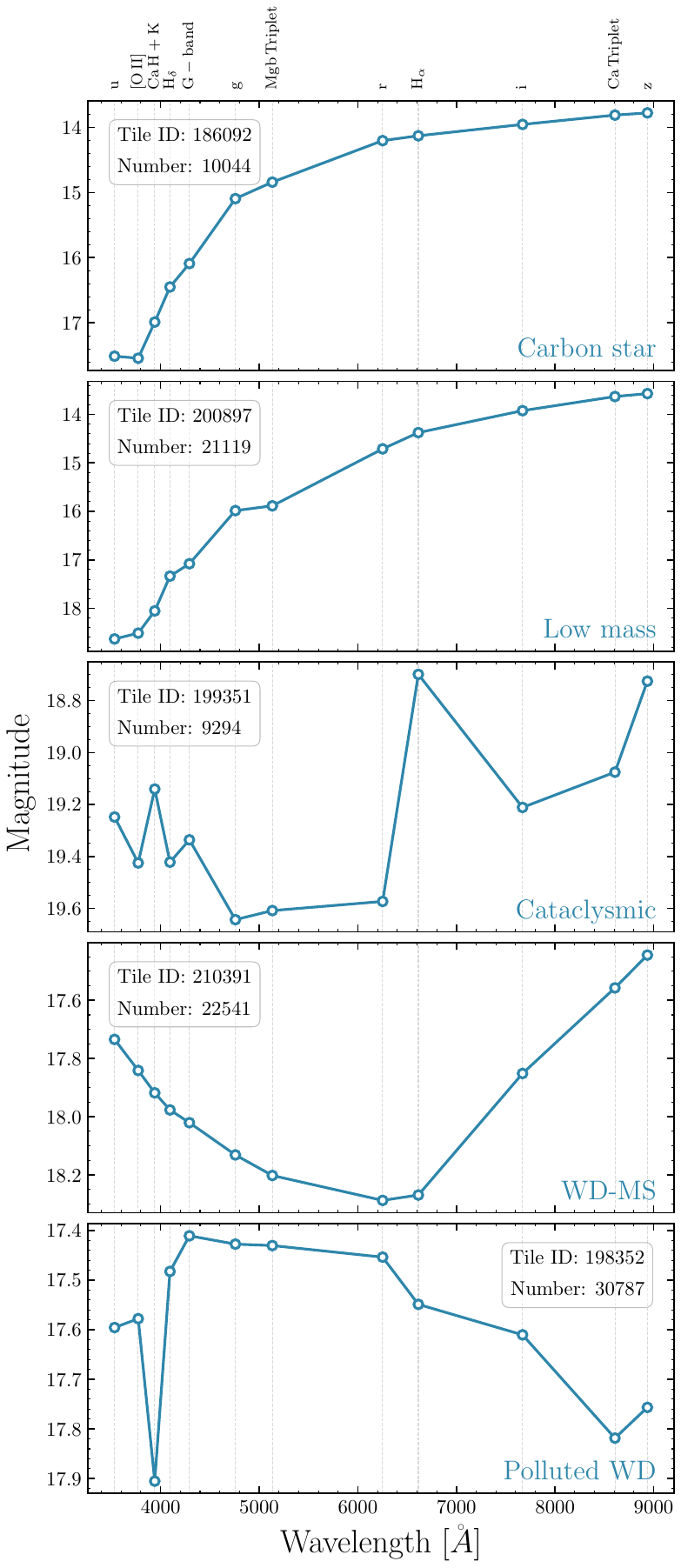}
\caption{Examples of the typical J-PLUS spectral energy distribution of the five populations described in this work. From top to bottom: carbon-rich star, low-mass star, cataclysmic variable, white-dwarf–main-sequence binary, and polluted white dwarf.}
\label{fig:sed_populations}
\end{figure}

We applied the \texttt{scikit-learn} implementation of t-SNE in \texttt{Python}\footnote{\url{https://scikit-learn.org/stable/modules/generated/sklearn.manifold.TSNE.html}} to 66 color indices constructed from the 12 J-PLUS magnitudes. We work here with colors, instead of magnitudes, because they describe the relative shape of the SED and are insensitive to a common brightness offset. This allows us to examine whether the outliers occupy distinct regions of the color space, and to identify the types of sources associated with the anomalous sample. The color indices were normalized to zero mean and unit variance using the \texttt{StandardScaler()} function, and were used as input to the t-SNE. This normalization is important because the t-SNE embedding is based on similarities derived from distances in the high-dimensional feature space. Standardization places the color indices on a common variance scale, reducing the possibility that features with larger variances disproportionately influence the distance calculations and the resulting embedding.

The t-SNE algorithm has many hyperparameters, but the results are mainly driven by the perplexity and, to a lesser extent, by the number of iterations. Except for these two hyperparameters, all the others were left at their default values in the implementation. We tested perplexity values in the range 20–50, and number of iterations between 2000 and 5000. While minor variations in clump distribution and density were observed in the resulting 2D maps, the overall results remained stable across these parameters ranges. In the end, we adopted a perplexity of 50 with 5000 iterations, as these values produced maps with more clearly defined over-densities. The left panel of Fig.~\ref{fig:tsne_map} shows the t-SNE map colored by the reconstruction error. Over-densities are clear, and some regions are dominated by very large reconstruction errors (in red).

To determine whether the members of the different stellar clumps appearing in the t-SNE map correspond to previously classified objects, we cross-matched our outliers population with the SIMBAD database, identifying $\sim 21,000$ classified sources in common. To ensure a reliable classification, we adopted the SIMBAD primary classification provided by the \mytt{main_type} label and we ignored sources generically flagged as stars, since we were interested only in well-defined populations. These cuts resulted in a final sample of $\sim 16,000$ sources with a SIMBAD classification type. Based on this information, we highlight in the left panel of Fig.~\ref{fig:tsne_map} the over-densities corresponding to four stellar populations: low-mass stars, white dwarfs, hot subdwarfs, and evolved stellar populations, as well as an over-density corresponding to extragalactic sources. 

The right panel of Fig.~\ref{fig:tsne_map} shows the same t-SNE map, but colored by the probability of the object being a star. The probability was obtained with the neural network–based \texttt{LeMoNNADE} software \citep[see][]{2026arXiv260706662H}, which performs star-galaxy separation based on SEDs. This probability flag, \texttt{p\_star}, is much more accurate than the \texttt{SExtractor} flag initially used in this work, but it was not available at the time when we started our study. We note that most sources in our outliers population are actual stars (in red), but there is a well-clustered group at the bottom of the t-SNE map that corresponds to extragalactic sources (in blue). This should be expected because the \texttt{SExtractor} classification is based only in morphology, thus several extragalactic point sources were missclassified as stars. 

We identified $\sim9800$ outlier sources with $\mathtt{p\_star} \leq 0.75$. Using photometric redshift information (photo-z) from \texttt{LeMoNNADE}, we found that these sources have redshift measurements ranging from 0 to 3.2, with a median of $z \approx 1.4$. However, only $\sim8500$ of these sources have $\mathtt{p\_star}<0.1$ and $z > 0.05$, which are considered reliable photo-z measurements according to \citet{2026arXiv260706662H}. We verified that the central 90\% of the extragalactic contaminant population that clustered in the t-SNE map displays photo-z in the range 0.2-2.3, thus confirming its extragalactic nature.

The \texttt{LeMoNNADE} classification helps to validate our methodology, because not only the extragalactic sources are correctly identified by the autoencoder, but they are also properly clustered in the t-SNE map, eventually allowing for their easy segregation.

\section{Populations in the \lowercase{t}-SNE Space} \label{sec:five}

Since our goal is to identify stellar sources suitable for spectroscopic follow-up, or for more detailed studies, we focused our analysis on five stellar classes that cluster prominently in specific regions of the t-SNE map: carbon-rich stars, low-mass stars, cataclysmic variable systems (CVs), white-dwarf–main-sequence binaries (WD-MS), and polluted white dwarfs. Classification validation was carried out through cross-matching with SIMBAD and external catalogs using a radius of 1.5 arcsec. Figure~\ref{fig:sed_populations} shows example SEDs of the five stellar classes analyzed here.

\begin{figure}[ht]
    \centering
    \includegraphics[width=1\linewidth]{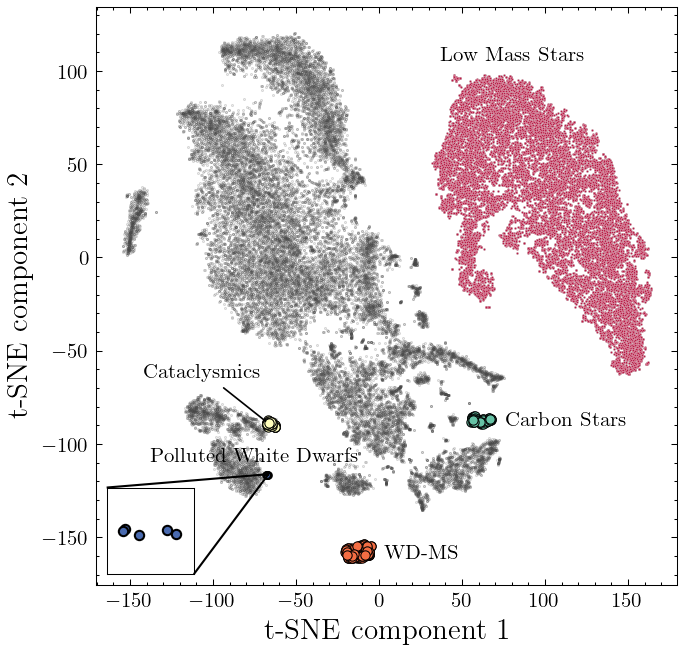}
    \caption{Recomputed two-dimensional t-SNE map for sources with $\mathtt{p\_star} > 0.75$, as provided by \mytt{LeMoNNADE}, displaying the populations described in this work. Carbon stars are shown in green, low-mass stars in pink, cataclysmic systems in yellow, white dwarf–main sequence (WD–MS) systems in orange, and polluted white dwarfs in purple in the zoomed-in region.}
    \label{fig:recomputed_tsne}
\end{figure}

To minimize the extragalactic contamination, we removed the sources with $\mathtt{p\_star} \leq 0.75$ and recomputed the t-SNE projection using only stellar sources. The resulting map is shown in Fig.~\ref{fig:recomputed_tsne}. Except for the low-mass stars and polluted white dwarf populations, the other clusters were extracted using the Density-Based Spatial Clustering of Applications with Noise (DBSCAN) algorithm \citep[DBSCAN;][]{1996kddm.conf..226E} applied to the t-SNE embedding. The t-SNE projection reveals well-defined density concentrations, making a density-based clustering approach appropriate for selecting these groups. The DBSCAN algorithm is basically controlled by two parameters: \texttt{eps}, that defines the neighborhood radius, and \texttt{min\_samples} which is the minimum number of points required to define a cluster. 

We adopted $\mathtt{eps = 2.5}$ and $\mathtt{min\_samples = 30}$, but we also tested robustness using $1.5\leq\mathtt{eps}\leq 4.0$ in steps of 0.5, and $20\leq\mathtt{min\_samples}\leq 50$ in steps of 10. At the adopted values, the carbon star, cataclysmic-variable, and WD-MS populations were each fully recovered in a single cluster, and remained $\geq95\%$ cohesive for all $\mathtt{min\_samples}$ with $\mathtt{eps}\geq2.5$. In parallel, we applied the HDBSCAN algorithm \citep{McInnes2017}, which does not require an $\mathtt{eps}$ parameter. Using $\mathtt{min\_cluster\_size=30}$, it recovered all the three populations as single clusters (102/102, 51/51, and 303/303 members, respectively), supporting the conclusion that these structures are robust to the clustering parameters rather than artifacts of the chosen boundaries.

\subsection{Carbon Stars}

Carbon stars are evolved objects characterized by a surface carbon abundance exceeding that of oxygen (C/O $>$ 1). This leads to the formation of prominent molecular bands, such as C$_2$, CN, and CH, that dominate their optical and infrared spectra \citep{Wallerstein1998, 2010JApA...31..177L}. These stars play an important role in stellar nucleosynthesis and in the chemical enrichment of the interstellar medium, particularly through the production of s-process elements. At low metallicities, carbon-enhanced metal-poor (CEMP) stars — commonly defined by [C/Fe] $\geq +1.0$ and [Fe/H] $\leq -1.0$ — constitute a significant fraction of the Galactic halo population \citep{Beers2005}. These stars are classified in subclasses based on their neutron-capture element abundances, including CEMP-s and CEMP-no \citep{Beers2005, 2007ApJ...655..492A, 2021A&A...649A..49G}. 

\begin{figure}[ht]
\centering
\includegraphics[width=\linewidth]{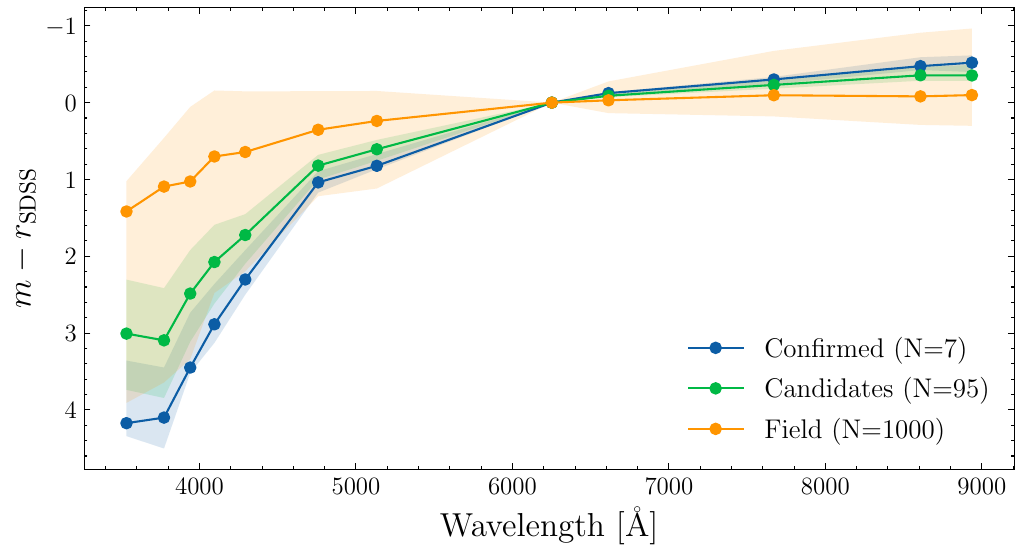}
\caption{Median normalized spectral energy distributions ($m-r_{\rm SDSS}$ versus wavelength) for the confirmed carbon stars (blue, $N=7$), the unconfirmed carbon star candidates (red, $N=95$), and a field sample of 1000 sources drawn from the rest of the anomalous population. Shaded regions mark the 16th--84th percentile range. The candidates closely track the red continuum of the confirmed carbon stars and are well separated from the field sample, consistent with their identification as carbon star candidates.}
\label{fig:carbon_sed_test}
\end{figure}

\begin{deluxetable}{rrrrrrrr}
\tabletypesize{\scriptsize}
\tablecaption{Candidate carbon-rich stars identified in J-PLUS, previously unreported in the literature. \label{tab:carbon_cand}}
\tablehead{
  \colhead{Number} & \colhead{Tile ID} & \colhead{$\alpha$ ($^\circ$)} & \colhead{$\delta$ ($^\circ$)} & \colhead{$g_{\rm SDSS}$} & \colhead{RUWE} & \colhead{$P_{\rm Star}$} & \colhead{Flag}
}
\startdata
20130 & 186309 & 207.464 & 26.092 & 17.50 & 0.94 & 0.9 & \\
13506 & 193399 & 154.430 & 40.265 & 15.73 & 3.36 & 1.0 & \\
50318 & 199373 & 279.222 & 50.645 & 14.83 & 0.99 & 1.0 & \\
28855 & 199411 & 105.583 & 52.452 & 17.45 & 0.95 & 1.0 & \\
22489 & 201308 & 108.250 & 56.727 & 15.38 & 0.96 & 1.0 & \tablenotemark{a} \\
29266 & 201871 & 266.153 & 56.644 & 18.37 & 1.04 & 1.0 & \\
11626 & 206563 & 338.233 & 11.855 & 14.91 & 1.09 & 1.0 & \\
4770  & 206719 & 333.620 & 13.518 & 15.60 & 1.14 & 1.0 & \\
24074 & 207355 & 333.734 & 18.557 & 15.46 & 1.05 & 1.0 & \\
35698 & 207511 & 333.878 & 19.453 & 15.78 & 1.05 & 1.0 & \\
17672 & 207871 & 333.289 & 22.889 & 15.24 & 0.99 & 1.0 & \\
21850 & 208029 & 31.853  & 22.597 & 16.58 & 1.03 & 1.0 & \\
29325 & 209299 & 350.024 & 30.839 & 16.15 & 0.97 & 1.0 & \\
29905 & 209381 & 34.721  & 30.638 & 15.58 & 1.18 & 1.0 & \\
32119 & 210154 & 27.621  & 33.412 & 15.98 & 1.04 & 1.0 & \\
8212  & 210185 & 32.668  & 34.256 & 16.69 & 1.91 & 1.0 & \\
10224 & 210413 & 28.385  & 35.643 & 15.75 & 1.09 & 1.0 & \\
57043 & 212923 & 276.735 & 31.478 & 13.93 & 1.02 & 1.0 & \\
75566 & 212948 & 284.055 & 32.425 & 14.50 & 1.09 & 1.0 & \\
\enddata
\tablenotetext{a}{APOGEE DR19 spectrum compatible with a carbon star.}
\end{deluxetable}

In this study, we identified a cluster of 102 sources (green dots in Fig.~\ref{fig:recomputed_tsne}), of which 21 ($\sim$21\%) have a SIMBAD literature classification. Only 7 of these are classified as carbon stars according to the \texttt{main\_type} flag. Other 5 sources with \texttt{main\_type} flag of chemically peculiar have \texttt{sp\_type} flag of CH, dC or carbon star candidate. Thus, 12 of the 21 sources with literature classifications ($\sim$57\%) are either confirmed carbon stars or consistent with the carbon star interpretation. The remaining 90 sources, for which no SIMBAD classification supports a carbon star identification, could be considered as new carbon star candidates based on their membership to this cluster. Nevertheless, several of those new sources have already been reported as carbon star candidates, specifically CEMP stars, in the LAMOST \citep{2025ApJS..277...30F} and \textit{Gaia} \citep{2023MNRAS.523.4049L} surveys.

Here, we assumed that only the seven cluster members with a confirmed carbon star main classification in SIMBAD were reliable, treating the remaining 95 cluster members as candidates. Then, we tested whether these 95 unconfirmed sources were photometrically consistent with carbon stars. The test consisted in comparing their median normalized J-PLUS 12-band SEDs (referenced to $r_{\rm SDSS}$) with those of the seven confirmed carbon stars and with a sample of 1000 field sources randomly drawn from the rest of the anomalous sample.

As shown in Fig.~\ref{fig:carbon_sed_test}, the seven confirmed carbon stars are separated from the field sources in SED space, with an RMS distance of 1.61 mag, consistent with their characteristically red colors and strong molecular absorption features. In contrast, the 95 unconfirmed candidates lie closer to the confirmed carbon star population (0.6 mag) than to the field sample (1 mag), by a factor of $\sim$1.7. Although the confirmed-carbon median is based on only seven objects, the unconfirmed candidates also cluster tightly with the confirmed carbon stars in t-SNE space.

Together, these results provide evidence that the 95 unconfirmed members are more consistent with carbon star candidates than with field contaminants. This is expected given that some J-PLUS filters, such as J0430, J0515, and J0861, sample spectral regions affected by strong molecular absorption bands, including the $\mathrm{C}_2$ Swan bands and CN, which are characteristic of carbon stars and produce flux deficits relative to the continuum.

Table~\ref{tab:carbon_cand} lists 19 carbon star candidates identified in this study that are not cataloged in SIMBAD, nor in the catalogs mentioned above. All sources have nearly a 100\% probability of being stars. With the exception of the source with Number 13506 and Tile ID 193399, all exhibit good astrometric solutions according to the Gaia RUWE parameter. One of these candidates has a spectrum available in the APOGEE DR19 catalog (2M07125989+5643385), and its spectral features are consistent with those of a carbon star. The full list of the 102 carbon star sources detected in J-PLUS is provided as a separate electronic table in machine-readable format.

\subsection{Low mass stars}\label{low_mass_stars}

Low-mass stars make up approximately 70\% of the stellar population in the Galaxy \citep{2010AJ....139.2679B}. Their low stellar masses enhance the detectability of terrestrial planets in the habitable zone, making the identification and observation of these stars of great interest \citep{2016PhR...663....1S}. In addition, these stars are valuable tracers of Galactic archaeology, as their long main-sequence lifetimes allow stars formed throughout the history of the Galaxy to remain observable today \citep{2024ARA&A..62..593H}.

The identification of this population as an outlier by the autoencoder was expected, given the low density of sources with $M_\mathrm{G} > 8.1$ in the training set, as shown in the color-magnitude diagram (Fig. \ref{fig:cmd_training_sample}). This population constitutes one of the largest clusters in the t-SNE map, as shown in Fig.~\ref{fig:recomputed_tsne}, located to the right and containing $\sim 15, 000$ members. The cross-match of this cluster with APOGEE and LAMOST yielded approximately 4000 sources in common, whose stellar parameters are consistent with low-mass stars.  

\subsection{Cataclysmic Variables}\label{cvs}
Close binary systems consisting of a white dwarf accreting matter from a late-type main-sequence companion that fills its Roche lobe are known as cataclysmic variables \citep{1995cvs..book.....W}. A more detailed discussion of the classification and physical processes operating in these complex systems is presented by \citet{Webb2023}, \citet{BelloniSchreiber2023}, and references therein.

We identified a cluster in the t-SNE map containing 51 sources (yellow dots in Fig.~\ref{fig:recomputed_tsne}), of which 44 ($\sim$86\%) have SIMBAD literature classifications. Of these, 39 are confirmed CVs and one is listed as a CV candidate. Thus, 40 of the 44 sources with literature classification ($\sim$91\%) have a CV classification. This yields a total of 12 CV candidates identified in this work, including 11 sources that have not been previously reported as CV candidates. They are listed in Table~\ref{tab:cv_cand}. All sources have a probability of being stars above 80\%. With the exception of the sources with Number 18998 (Tile ID 198451) and 33264 (Tile ID 211373), all exhibit good astrometric solutions according to the Gaia RUWE parameter. By visually inspecting the SEDs of these 12 candidates, we verified that seven of them show a peak in the J0660 band, covering the H$\alpha$ emission line, which is a well-known signature of accretion-related processes in cataclysmic variables \citep{1995cvs..book.....W}. The remaining sources, which do not show a clear J0660 excess, are flagged in Table~\ref{tab:cv_cand}. Cross-matching the full anomalous sample with SIMBAD yielded 15 additional sources classified as CVs; however, these do not appear well clustered in the t-SNE map. This is expected, since cataclysmic variables are complex systems, and their SEDs may not exhibit consistent patterns across the full spectral range, or they may be misclassified. The full list of the 66 CVs sources (51 clustered + 15 not clustered) detected in the J-PLUS is provided as a separate electronic table in machine-readable format.

\begin{deluxetable}{rrrrrrrr}
\tabletypesize{\scriptsize}
\tablecaption{Candidate cataclysmic variable systems identified in the J-PLUS. Only one star has been previously reported in the literature. \label{tab:cv_cand}}
\tablehead{
  \colhead{Number} & \colhead{Tile ID} & \colhead{$\alpha$ ($^\circ$)} & \colhead{$\delta$ ($^\circ$)} & \colhead{$g_{\rm SDSS}$} & \colhead{RUWE} & \colhead{$P_{\rm Star}$} & \colhead{Flag}
}
\startdata
18998 & 198451 & 216.168 & 49.499 & 16.20 & 2.19 & 0.9 & \tablenotemark{b}\\
9294  & 199351 & 275.125 & 51.747 & 19.64 & 1.03 & 1.0 & \\
37140 & 199951 & 265.154 & 52.195 & 17.84 & 1.06 & 0.8 & \\
16814 & 203238 & 275.485 & 69.580 & 17.22 & 1.06 & 0.9 & \tablenotemark{b}\\
4603  & 206323 & 1.177   & 9.301  & 16.88 & 1.06 & 1.0 & \tablenotemark{a}\\
40503 & 206707 & 332.204 & 12.362 & 17.41 & 1.00 & 1.0 & \tablenotemark{b}\\
26667 & 207188 & 332.120 & 17.105 & 19.15 & 0.99 & 1.0 & \\
2444  & 207211 & 336.135 & 17.780 & 18.02 & 1.03 & 0.8 & \\
11610 & 207224 & 337.539 & 17.522 & 17.81 & 0.96 & 1.0 & \tablenotemark{b}\\
51768 & 209442 & 334.616 & 31.855 & 17.93 & 1.03 & 1.0 & \tablenotemark{b}\\
33264 & 211373 & 10.911  & 41.756 & 18.62 & 2.18 & 1.0 & \\
19840 & 212371 & 286.607 & 28.491 & 17.68 & 1.34 & 1.0 & \\
\enddata
\tablenotetext{a}{\footnotesize Classified as a CV candidate in SIMBAD.}
\tablenotetext{b}{\footnotesize SED does not show a clear signature at the J0660 band.}
\end{deluxetable}

\subsection{White-dwarf–main-sequence binaries}
The search and the identification of white-dwarf–main-sequence (WD–MS) binary systems is important because these systems place constraints on binary evolution and stellar evolution \citep{2016MNRAS.458.3808R}. Recently, \citet{2025ApJS..279...47L} report 30,000 WD-MS binary candidates using \textit{Gaia} DR3's low-resolution BP/RP (XP) spectra.

In the t-SNE map, we identified a cluster containing 303 sources (orange dots in Fig.~\ref{fig:recomputed_tsne}). According to the SIMBAD \texttt{main\_type} flag, 40 sources are classified as white dwarfs (WDs), while 52 are classified as multiple systems. Since the latter classification does not specify the nature of these systems, we additionally examined their spectroscopic type information (\texttt{sp\_type}). Among the 118 sources with available classifications, 86 ($\sim$73\%) have a WD--MS classification. Furthermore, in the color-magnitude diagram, the majority of the 303 sources lie between the white dwarf and main-sequence stellar populations, further supporting the interpretation of the cluster as a population of WD--MS candidates.

To further assess the nature of this cluster, we crossmatched the 303 sources with several external catalogs. We found 56 sources in common with the WD--MS Gaia candidates of \citet{2025ApJS..279...47L}, 4 sources in common with the J-PLUS DR3 white dwarf catalog \citep{2025A&A...701A.273L} that are classified as WD--MS candidates, and 47 sources in common with the catalog of \citet{2025yCat..36990153R}. These independent catalog matches provide additional support for the WD--MS interpretation of the cluster. Excluding sources already classified in SIMBAD, as well as those previously identified as WD--MS candidates in the aforementioned catalogs, we remained with 170 new WD--MS candidates. Table~\ref{tab:wdms_stars} lists a sample of these candidates, while the complete catalog of the 303 sources in the cluster is provided as a supplementary electronic table in machine-readable format. The first 170 entries in that table correspond to the WD--MS candidates identified in this work.

\begin{deluxetable}{rrrrrrr}
\tabletypesize{\scriptsize}
\tablecaption{Sample of White-Dwarf-Main-Sequence systems identified in the J-PLUS.
\label{tab:wdms_stars}}
\tablehead{
  \colhead{Number} & \colhead{Tile ID} & \colhead{$\alpha$ ($^\circ$)} & \colhead{$\delta$ ($^\circ$)} & \colhead{$g_{\rm SDSS}$} & \colhead{RUWE} & \colhead{$P_{\rm Star}$}
}
\startdata
32540 & 185315 & 107.371 & 20.576 & 18.98 & 1.35 & 0.9 \\
30794 & 185404 & 110.703 & 20.501 & 18.21 & 1.30 & 1.0 \\
25145 & 185438 & 106.391 & 20.667 & 18.47 & 2.50 & 1.0 \\
10095 & 185672 & 142.259 & 55.800 & 17.90 & \nodata & 1.0 \\
11364 & 185969 & 109.779 & 25.214 & 18.01 & 4.31 & 1.0 \\
\enddata
\end{deluxetable}

\subsection{Polluted white dwarfs}
Strong metal features such as magnesium, iron, and calcium are not expected in white dwarf atmospheres, as these elements rapidly sink due to the star’s high gravity. Their detection indicates pollution from recent accretion of rocky planetary debris, making the atmospheric chemical composition a proxy for studying planetary systems \citep{2018MNRAS.479.3814H}. Recently, non-supervised machine learning techniques applied to low-resolution \textit{Gaia} spectra have led to the identification of several polluted white dwarf candidates \citep{2024ApJ...970..181K, 2025RASTI...4af044B}.

We detected a small and tight cluster in the t-SNE map containing five potential polluted white dwarfs (purple dots in the zoomed-in Fig.~\ref{fig:recomputed_tsne}). Visual inspection of their SEDs revealed a strong drop in the J0395 band, which covers the prominent Ca II absorption feature, a key signature of polluted white dwarfs \citep{2024A&A...691A.211L}, with the depth of this feature depending on the degree of pollution. A crossmatch of this cluster with the Planetary Enriched White Dwarf Database \citep[PEWDD;][]{2024A&A...691A.352W} yielded three sources in common that are confirmed polluted white-dwarfs. The remaining two candidates was already reported in \citep{2025A&A...701A.273L}.
The crossmatch of our entire anomalous sample with the PEWDD yielded 24 additional sources. However, these sources are not clearly clustered in the t-SNE map, mainly because their Ca\,\textsc{ii} absorption feature is not sufficiently prominent. 



\section{Discussion and conclusions}\label{sec:six}

Identifying particular stellar population candidates is relevant for subsequent spectroscopic follow-up or other studies. This work demonstrates that unsupervised machine-learning techniques, like autoencoders, applied to photometric data are effective tools for the classification and selection of different stellar populations candidates.
In particular, this proves to be quite efficient when the photometry comes from a system of broad- and narrow-band filters, with the narrow-band filters strategically placed in key spectral regions, as is the case of the J-PLUS system. We recall, however, that the application requires to be complemented with reliable external information, either from the literature or from other astronomical surveys.

In our application, both the number of outliers identified by the autoencoder and the total number of reported sources are affected by incompleteness with respect to the full parent catalog. This is due to several factors that we discuss in the following.

\subsection{Training sample selection}

Detection of outliers by the autoencoder strongly depends on the populations present in the training sample. Sources with magnitudes outside the range covered by the training sample may be identified as anomalous by the autoencoder, due to limited generalization beyond the training domain. We verified that the selection cuts applied to build the training sample (Sect.~\ref{sec:train_val_test_samples}) did not alter its magnitude range, nor affected its composition. The \texttt{RUWE} cut was responsible for removing the largest number of objects: $\sim 8$\% of the initial sample; the majority of these were binary systems, though some binaries might still pass the \texttt{RUWE} cut. Such exclusion increased the probability that binaries be detected as anomalous objects, given their small relative presence in the training sample. Nevertheless, this did not significantly affect the autoencoder's representation of the majority populations.

\subsection{Autoencoder architecture}\label{sec:auto-arch}

High-capacity autoencoders can generalize well to minority populations, reducing their detection as anomalies \citep[and references therein]{zavrtanik2021107706, Gong_2019, Bouman_2025}. To mitigate this, various techniques have been proposed, including regularization methods \citep{10.1145/3097983.3098052, zavrtanik2021107706} and latent space constraints \citep{Astrid_2024}. We adopted a combined strategy to constrain the model's representational capacity and reduce overfitting. During hyperparameter tuning, we varied the latent-space dimensionality between 2 and 6 units and applied L1 and L2 regularization to the weights of the encoder, bottleneck, and decoder hidden layers. The optimal configuration selected a five-dimensional latent space for the final model. The L1 and L2 terms constrained the learned weights by promoting sparsity and penalizing large weight magnitudes, respectively, thereby limiting the model's ability to memorize individual sources. Combined with the reduced latent dimensionality, these constraints encouraged the autoencoder to capture the dominant patterns in the stellar SEDs.

\subsection{Outlier threshold}

The number of sources classified as outliers or anomalous, after applying the autoencoder, depends on the chosen reconstruction-error threshold, which can affect the populations identified in the t-SNE plane. The choice depends on the objective of the study, and higher threshold values may be appropriate if the aim is to identify only the most extreme or potentially atypical objects. In this work, we adopted a 1$\sigma$ threshold to identify sources as outliers. While this criterion is lower than thresholds commonly used in the literature for outlier detection (typically 2$\sigma$ or 3$\sigma$), it represents a reasonable choice to avoid missing potentially interesting objects that would be excluded at higher threshold values (see Sect.~\ref{subsec:app}).

\subsection{Observational errors}\label{sec:obs-errors}

\begin{figure}[ht]
\centering
\includegraphics[width=\linewidth]{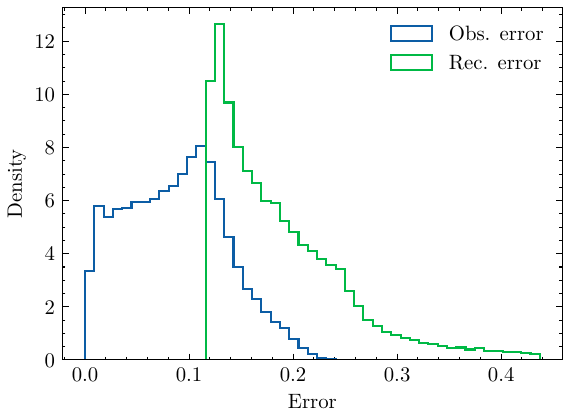}
\caption{Comparison of the observational error (blue) and the reconstruction error (green) distributions. The normalized histograms show the probability densities of each error, with the x-axis restricted to the 99th percentile of the combined error distribution to reduce the influence of extreme outliers.}
\label{fig:error_distribution}
\end{figure}

The photometric errors of the input data (i.e. the observational errors in each individual filter) were not taken into account in the training of the autoencoder. In principle, the autoencoder is not designed to account for intrinsic errors in the training instances. In order to determine whether the photometric errors may affect the anomalous sample selection, we computed, for each anomalous source, its observational error:
\begin{equation}
    E_{\mathrm{obs}}=2\left[\frac{1}{N}\sum_{j=1}^{N}\left(\delta m_{j}^{\mathrm{obs}}\right)^{2}\right]^{1/2} \label{eq:obs_err}
\end{equation}
where $N = 12$, and $\delta m_{j}^\mathrm{obs}$ is the photometric uncertainty in each filter, assuming that the observed magnitude in a filter is given by $m_{j}^\mathrm{obs} \pm \delta m_{j}^\mathrm{obs}$. These observational errors do not necessarily follow a decreasing distribution, but instead exhibit a strongly right-skewed distribution (blue histogram in Fig.~\ref{fig:error_distribution}). 

The distribution of $\sqrt{E_{\mathrm{rec}}}$ (green histogram in Fig.~\ref{fig:error_distribution}) can be compared using the corresponding quantiles, $Q_n$. For the reconstruction errors of the anomalous sample, we obtained a minimum of 0.12, a median ($Q_{0.5}$) of 0.17, and a dispersion ($Q_{0.75}-Q_{0.25}$) of 0.09. On the other hand, for the observational errors, the median is 0.09, the dispersion is 0.07, and the maximum is 0.24. This indicates that, in most cases, the observational errors are smaller than the reconstruction errors as shown in Fig.~\ref{fig:error_distribution}, and they are unlikely to significantly affect the composition of the anomalous sample, except for the sources very close to the selection threshold.

\subsection{t-SNE hyperparameters}

The t-SNE hyperparameters influence the configuration in lower dimensions, but the final result is primarily defined by the perplexity parameter. The perplexity parameter balances attention between local and global aspects of the data by controlling the effective number of nearest neighbors considered in the embedding \citep{vanDerMaaten2008}. In this work, the perplexity was tuned using the SIMBAD classifications available for some sources in the
overdensity regions of the t-SNE map, with the aim of favoring the grouping of sources with similar SIMBAD classifications, together with visual inspection of the SEDs.

It is important to note that t-SNE is primarily a visualization tool, and may produce apparent cluster structures that do not necessarily reflect true clustering in the original data \citep{2025RASTI...4af044B, 2025arXiv251007746B}. Therefore, groups in the embedded space should be always interpreted with caution, and validated using additional information. In our case, besides the cross-match with other databases, the visual inspection of the spectral energy distributions proved relevant to confirm or reject classification. Therefore, while the cross-match approach fails in clumpy areas without known sources in the literature, inspecting the SEDs can help in the identification of such groups.

Actually, we may expect that some sources, that could be considered as anomalous, were not actually detected by our method because their reconstruction errors are slightly lower than the applied threshold. A more detailed discussion on these limitations can be found in \cite{2025AJ....169..332Q}.

\subsection{Conclusion} 

Our methodology applied to the J-PLUS DR4 catalog allows us to potentially identify carbon-rich stars (most of them CH and CEMP stars), cataclysmic variables, polluted white-dwarfs, white-dwarfs-main-sequence binaries, and $\sim 15,000$ low-mass stars. Of these, 19 carbon-rich star candidates, 11 cataclysmic variable candidates, and 170 white-dwarfs-main-sequence binary candidates are previously unreported in SIMBAD and other catalogs.

Recognizing the importance of medium- and high-resolution spectroscopic follow-up to validate our findings, we are preparing a proposal to request telescope time to observe the brightest carbon-rich star candidates identified in this work. Notably, observations of 5 carbon-rich candidates identified by \cite{2025AJ....169..332Q}, applying the same methodology presented here to the S-PLUS catalog, confirm that they are, in fact, carbon rich stars (Quispe-Huaynasi et al., in preparation).

\vspace{1cm}
\noindent F.Q.H. acknowledges fellowship support from the Espirito Santo State Science Foundation (FAPES) and from the Brazilian National Council of Research (CNPq — PCI Program). F.R. acknowledges support from CNPq grant 312429/2023-1. V.L.T. acknowledges support from the CNPq through the Postdoctoral Junior (PDJ) fellowship, process 152242/2024-4. The work of V.M.P. is supported by NOIRLab, which is managed by the Association of Universities for Research in Astronomy (AURA) under a cooperative agreement with the U.S. National Science Foundation. VM thanks CNPq (Brazil), CAPES (Brazil) and FAPES (Brazil) for partial financial support. S.D. acknowledges CNPq/MCTI for grant 306859/2022-0 and FAPERJ for grant 210.688/2024.

Based on observations made with the JAST/T80 telescope at the Observatorio Astrofísico de Javalambre (OAJ), in Teruel, owned, managed, and operated by the Centro de Estudios de Física del Cosmos de Aragón (CEFCA). We acknowledge the OAJ Data Processing and Archiving Unit (UPAD) for reducing and calibrating the OAJ data used in this work. 

Funding for the J-PLUS Project has been provided by the Governments of Spain and Arag\'on through the Fondo de Inversiones de Teruel; the Aragonese Government through the Research Groups E96, E103, E16\_17R, E16\_20R, and E16\_23R;  grants PID2024-162229NB-I00, PID2024-155572NB-C21, and PID2024-155572NB-C22 (MICIU/AEI/10.13039/501100011033 and ERDF/EU); the Spanish Ministry of Science and Innovation (MCIN/AEI/10.13039/501100011033 y FEDER, Una manera de hacer Europa) with grants PID2021-124918NB-C41, PID2021-124918NB-C42, PID2021-124918NA-C43, and PID2021-124918NB-C44; the Spanish Ministry of Science, Innovation and Universities (MCIU/AEI/FEDER, UE) with grants PGC2018-097585-B-C21 and PGC2018-097585-B-C22; the Spanish Ministry of Economy and Competitiveness (MINECO) under AYA2015-66211-C2-1-P, AYA2015-66211-C2-2, AYA2012-30789, and ICTS-2009-14; and European FEDER funding (FCDD10-4E-867, FCDD13-4E-2685). The Brazilian agencies FINEP, FAPESP, and the National Observatory of Brazil have also contributed to this project.

This work used of data from the European Space Agency (ESA) mission \textit{Gaia} (\url{https://www.cosmos.esa.int/gaia}), processed by the \textit{Gaia} Data Processing and Analysis Consortium (DPAC, \url{https://www.cosmos.esa.int/web/gaia/dpac/consortium}). Funding for the DPAC has been provided by national institutions, in particular the institutions participating in the \textit{Gaia} Multilateral Agreement.


\bibliography{sample631}{}
\bibliographystyle{aasjournal}

\end{document}